\documentclass[5p]{elsarticle}
\usepackage{graphicx}
 \usepackage{xcolor}
 \newcommand{\bc}{\begin{center}}
 \newcommand{\ec}{\end{center}}
 \newcommand{\be}{\begin{equation}}
 \newcommand{\ee}{\end{equation}}
 \newcommand{\bea}{\begin{eqnarray}}
 \newcommand{\eea}{\end{eqnarray}}

 \newcommand{\bt}{\begin{tabbing}}
 \newcommand{\et}{\end{tabbing}}

\begin{document}

\title{Atomic, electronic, and superconducting properties of  Zr$_2$Ir compound}

\author{C.      Tayran$^{1}$ and      M.      ~\c{C}akmak$^{2,3}$}
\address{$^{1}$Department of  Physics, Gazi   University,    06500   Ankara,    Turkey}

\address{$^{2}$Department  of Photonics, Faculty of Applied Sciences, Gazi
  University, 06500 Ankara, Turkey}

\address{$^{3}$Photonics   Application  and   Research  Center,   Gazi University, 06500 Ankara, Turkey}

\begin{abstract}
We have investigated  the structural, electronic, mechanical, phononic, and superconducting properties of the Zr$_2$Ir compound with a body-centered tetragonal crystal structure using first-principles calculations. Our analysis reveals that the Zr$_2$Ir compound shows mechanical and dynamically stable  by using with and without spin-orbit coupling (SOC) effect.   After calculating some properties  such as elastic constants, Bulk modulus, Young's modulus, Poisson ratio, Debye temperature, and  sound velocity, we found that Zr$_2$Ir is ductile. When the elastic constants $C_{11}$ and $C_{33}$ are compared, it is determined that the situation changes in the opposite direction under the effect of SOC, that is, more compressibility along the x-axis turns into the z-axis. Here, the electronic band structure and intensity of the states calculated for the compound show a metallic character. The superconducting critical temperature ($T_c$) and electron-phonon coupling constant ($\lambda$) were found to be 7.50 K and 0.93 without SOC and 7.62 K and 0.96 with SOC, respectively.
We determined that although the Zr$_2$Ir compound has a strong electron-phonon coupling regime, the inclusion of SOC slightly reduces its critical temperature and electron-phonon coupling constant.
\end{abstract}

\maketitle

\section{Introduction}\label{intro}

High temperature-resistant compounds are leading materials for industry and research. For instance, Ir-based intermetallic compounds are considered potential high temperature structural materials~\cite{Mit-99, Wu-17, Cor-03, Che-03, Ran-06}, with high melting temperatures and superior oxidation resistance~\cite{Ran-06}.
Also, high temperature alloys are preferred in turbine engines to increase power output efficiency. Some alloys also show excellent mechanical properties at elevated temperatures. For this reason, Ir-based alloys are of great interest~\cite{Mit-05, Sha-06}.  In fact, as many experimental and theoretical studies have shown, the superior properties of the material are closely related to the intrinsic properties of Ir-based superalloys containing Zr phases. Understanding the intrinsic properties of the intermetallic compound, such as mechanical or electronic properties, is essential for the design of composite materials~\cite{Zhan-18}.

While Kuprina et al.~\cite{Kup-74} and  Eremenko et al.~\cite{Ere-74, Ere-78, Erem-78} previously worked on the phase diagram of the Ir–Zr system, Okamoto et al.~\cite{Oka-92} studied structural, mechanical, thermodynamic, and dislocation properties of Ir-Zr binary system (e.g. Ir$_3$Zr,  Ir$_2$Zr, IrZr, Ir$_3$Zr$_5$, IrZr$_2$, and IrZr$_3$). In addition to the phase studies of Eremenko et al.~\cite{Ere-74, Ere-78, Erem-78} and Kuprina et al.~\cite{Kup-74}, Eremenko et al.~\cite{Ere-80} later using high-temperature X-ray and thermal analysis experimental studies revealed that the IrZr$_2$ structure is tetragonal Al$_2$Cu-type (I4/mcm, Z = 4). On the theoretical side, Wu et al.~\cite{Wu-17} presented the structural, mechanical, electronic, and thermodynamic properties of the Ir–Zr compounds (Ir, Ir$_3$Zr, Ir$_2$Zr, IrZr, B2-IrZr, Ir$_3$Zr$_5$, IrZr$_2$, IrZr$_3$ and $\alpha$-Zr) using density functional theory to ensure useful and practical information for further experimental and theoretical investigations on the Ir–Zr alloys. Furthermore, Zr$_2$Ir is also superconducting material and the critical temperature of Zr$_2$Ir was found to be $T_c$ 7.3 K and 7.5 K in different studies~\cite{Mc-71, Fis-74, Fag-06, Kas-21}. Recently, Mandal et al.~\cite{Man-21} presented the superconductivity properties of  nonsymmorphic Zr$_2$Ir compound. In that study, superconductivity properties were investigated using magnetic susceptibility, electrical resistivity, heat capacity, and muon-spin rotation/relaxation ($\mu$SR) techniques. Their results indicate that Zr$_2$Ir can be described in the superconducting state as fully gapped s-wave order parameters and conserved time-reverse symmetry. They estimated transition temperature $T_c$=7.4 K of Zr$_2$Ir is bulk type-II superconductivity. However, a more comprehensive study has not yet been done for these features, and the deficiencies in the meaning of the SOC effect have not yet been answered.

In this paper, we have investigated the electronic, mechanical, and superconducting properties of Zr$_2$Ir using the first-principles study. The band structure, total and partial density of states and electronic charge density were systematically studied. The various mechanical properties such as the bulk modulus, shear modulus, Young's modulus, Poisson's ratio, Pugh ratio, and Cauchy pressure were  calculated. The inclusion of SOC has an effect on mechanical properties such as  it is greater compressibility along the z-axis.  The critical temperature $T_c$ was found  to be 7.50 K with SOC and 7.62 K without SOC which is in good agreement with the  experimental values (7.30 K, 7.40 K, 7.50 K). The inclusion of SOC slightly reduces the critical temperature.

\section{Method}

\begin{figure}
\centering
\bc
\includegraphics[width=9.5 cm]{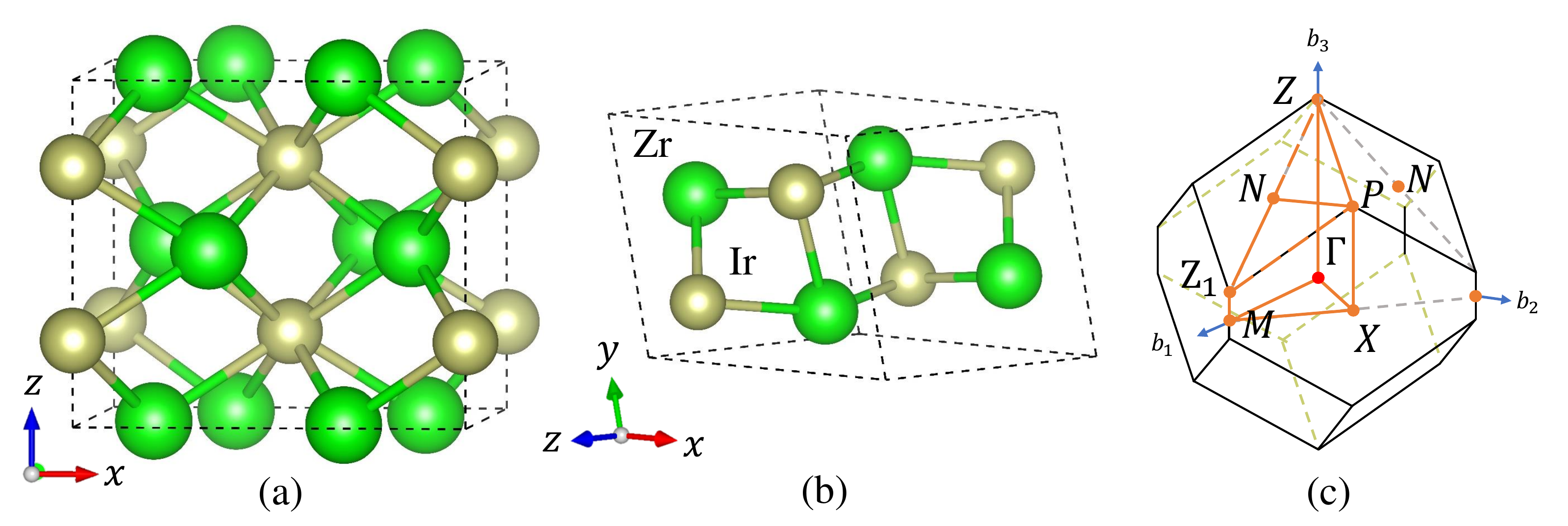}
\ec
\caption{\label{fig:fig-1} The atomic structures for (a) conventional, (b) primitive cell of Zr$_{2}$Ir, and
(c) schematic drawing of Brillouin zone (BZ) with high symmetry points. Green
and gold circles indicate Zr and Ir atoms, respectively.}
\end{figure}

Our calculations were performed within density functional theory (DFT) implemented in the Quantum Espresso package~\cite{Gia-09, Gia-17}  using fully relativistic (consider SOC) and scalar relativistic (ignore SOC) projector augmented wave (PAW)~\cite{Blo-94} data sets to describe the interaction between electron and ion. The exchange-correlation potential was based on the Generalized gradient approximation (GGA)  with the Perdew-Burke-Ernzerhof (PBE)~\cite{Per-96}.  The  tested 60 Ry and 480 Ry were also used for the cutoff energy and the charge density, respectively. The Broyden--Fletcher--Goldfarb--Shanno (BFGS) algorithm~\cite{Fis-92} was utilized to optimize the geometry. The total energy convergence criteria of 10$^{-6}$ Ry and the force convergence criteria of 10$^{-5}$  Ry/Bohr were considered. We  also adopted  the   Methfessel  Paxton broadening technique~\cite{Met-89}.  The Brillouin region was defined using Monkhorst-Pack~\cite{Mon-76} 24$\times$24$\times$24 $\textbf{\textit{k}}$--point mesh. The valence electrons of Zr and Ir are chosen as 4$s$$^{2}$4$p$$^{6}$5$s$$^{2}$4$d$$^{2}$ and 6$s$$^{2}$5$d$$^{7}$, respectively. Furthermore, the mechanical  properties were  computed using the thermo$\_$pw code~\cite{Cor-16,  qe}. In phonon calculations, the phonon spectrum of Zr$_{2}$Ir compound was calculated using density functional perturbation theory in the linear response  approach~\cite{Gia-09, Gia-17} and 4$\times$4$\times$4 uniform grid of q-points was taken for dynamical matrices. The electron-phonon matrix elements were calculated using a combination of the linear  response theory~\cite{Gia-09,  Gia-17} and the  Migdal-–Eliashberg theory~\cite{Mig-58,  Eli-60} on dense k- and q-grids of 24$\times$24$\times$24. To  estimate the
superconducting  transition temperature  ($T_c$),  Mc--Millan’s
equation  modified  by  Allen  and  Dynes was employed~\cite{Mil-68,  All-75a, All-75b}. For the Fermi surface 48$\times$48$\times$48 $\textbf{\textit{k}}$--point  grid was evaluated. VESTA~\cite{Mom-11} and XCrySDen~\cite{Kok-99} software programs were also used for plotting atomic structure, charge density, and Fermi surfaces respectively.

\section{Results and Discussion}

\subsection{Structural and electronic properties}

\begin{figure}
\centering
\bc
\includegraphics[width=8.5cm]{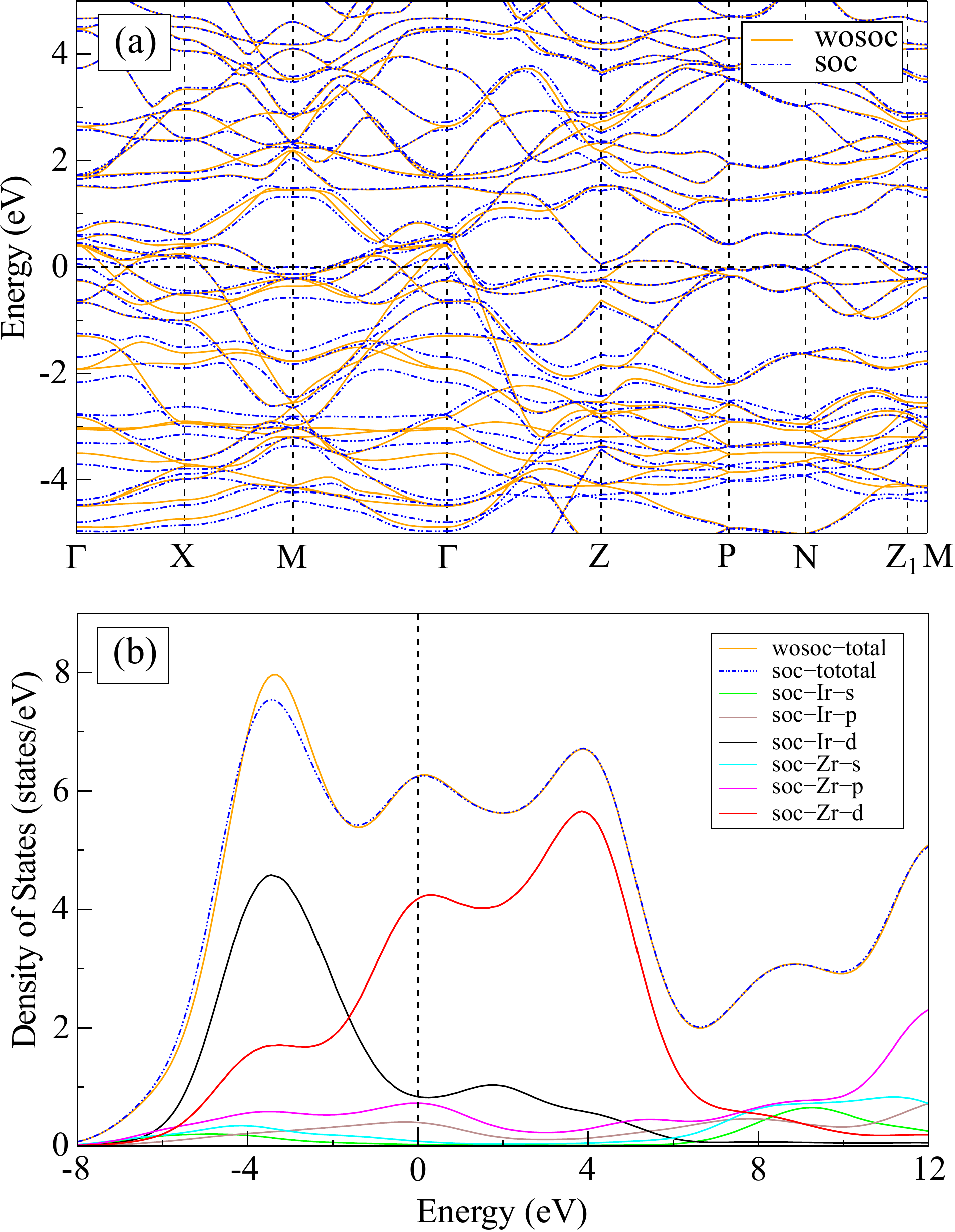}
\ec
\caption{\label{fig:fig-2} The calculated (a) electronic band structure along high symmetry directions in the BZ and (b) total and partial electronic density of states of Zr$_{2}$Ir with and without SOC, respectively.}
\end{figure}

The Zr$_{2}$Ir compound occurs in a body-centered tetragonal crystal structure with I4/mcm (140) space group~\cite{Ere-80}. There are 2 Ir and 4 Zr atoms in the primitive cell. The conventional and primitive cells of the Zr$_{2}$Ir compound are shown in Figs. \ref{fig:fig-1}(a) and (b).  We have calculated equilibrium structural key parameters such as lattice parameters and primitive cell volume for the Zr$_{2}$Ir compound. The lattice parameters are found to be value for $a$=6.541 {\AA}, $c$= 5.748 {\AA}  and 122.963 {\AA$^3$} as listed in Table \ref{tab:table-1}. The calculated results are slightly larger than the experimental values~\cite{Oka-92, Mc-71, Man-21}, but they are fairly consistent with other theoretical results~\cite{Wu-17, Jai-13} as listed in  Table \ref{tab:table-1}. The calculated distance between Zr and Ir atoms is also found to be 2.821 {\AA} for the Zr$_{2}$Ir structure.

After the structural description, we have calculated the electronic band structure at $\Gamma$-$X$-$M$-$\Gamma$-$Z$-$P$-$N$-$Z_1$-$M$ high symmetry points (c.f. in Fig. \ref{fig:fig-1}(c)) in the Brillouin zone (BZ) within the energy range from -5 to +5 eV. Fig. \ref{fig:fig-2}(a) shows the calculated electronic band structure with and without SOC,  which exhibits a metallic character due to the  overlap of the valence and conduction bands at the Fermi level. Indeed, the calculated band structure is in good agreement with the previously reported theoretical work~\cite{Jai-13}. When we further analyze the band structure at some high symmetry points such as $\Gamma$, $Z$, and $M$, we can clearly see that the SOC effect predominates due to the mass difference between the heaviest Ir atom and the lightest Zr atom.
The band splits seen in Fig.\ref{fig:fig-2}(a) are more intense at $\Gamma$ at about -2 eV below the Fermi level. To understand the electronic behavior of Zr$_{2}$Ir, the total and partial density of  states labeled TDOS and PDOS are shown in Fig.\ref{fig:fig-2}(b) with and without SOC. Our calculated TDOS and PDOS are consistent with existing theoretical studies~\cite{Wu-17, Jai-13}. It is clear that the main contributions to the total   density of  states come from the 4$d$ orbital of Zr and the 5$d$ orbital of Ir atoms as seen in PDOS. These orbitals contribute to the formation of Ir–Zr metallic bonds, revealing hybridized $d$-$d$ states~\cite{Wu-17}. In addition, at the Fermi level, the $d$ orbital of the Zr atom is most dominant, followed by the $d$-orbital of the Ir atom and the $p$ orbital of the Ir and Zr atoms. Most of the contributions above and below the Fermi level come from the $d$ orbital of the Zr and Ir atoms, respectively. Below the Fermi energy, the $d$ orbitals from the Zr and  Ir atoms intersect at an energy close to -2 eV.

\begin{table}
\begin{center}
\caption{\label{tab:table-1} The calculated
lattice parameters $a$ and $c$ (in \AA), volume $V$ (in \AA$^3$), distance between Zr and Ir atoms (in \AA ) and with the available theoretical and experimental values.}
\begin{tabular}{c|cccc}
\hline
                 & \textit{a}  &  \textit{c}  &   \textit{d$_{Zr-Ir}$} &  \textit{V}   \\\hline
This work           &  6.541          &  5.748      &    2.821              & 122.963 \\
Exp.~\cite{Oka-92}                 &  6.510          & 5.562      &                       & \\
Exp.~\cite{Mc-71}   &  6.508         & 5.721      &                       & \\
Exp.~\cite{Man-21}                &  6.516          & 5.661       &                       & 120.702\\
Theo.~\cite{Wu-17}              &  6.529          & 5.757       &                       & \\
Theo.~\cite{Jai-13}               &  6.543          & 5.779       &                       & 123.710
 \\\hline
\end{tabular}
\end{center}
\end{table}

\begin{figure}
\centering
\bc
\includegraphics[width=9cm]{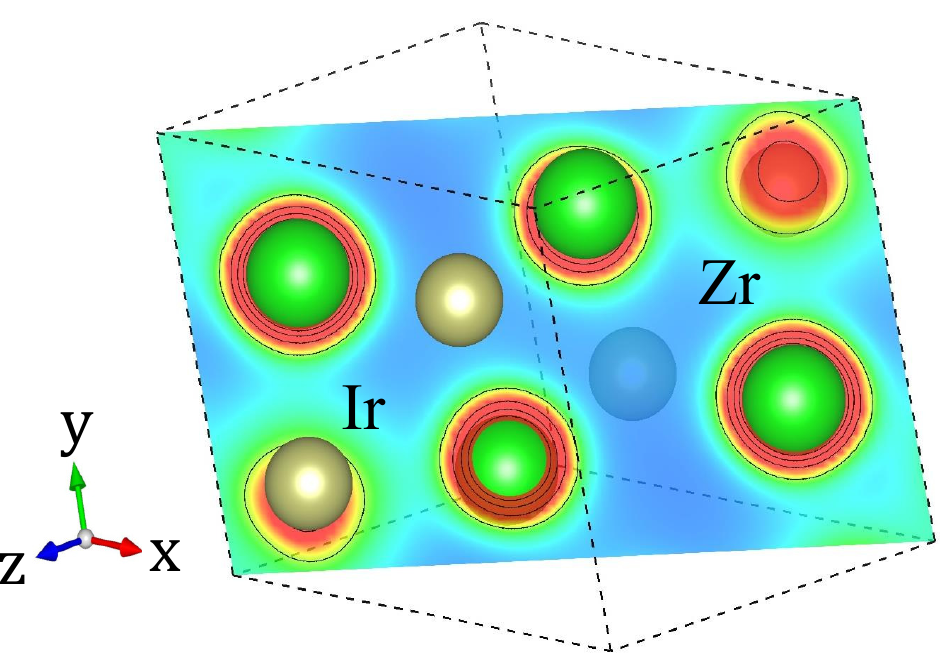}
\ec
\caption{\label{fig:fig-3} The calculated electronic total charge density  plots  for a  plane cutting through Zr-Ir atoms.}
\end{figure}

\begin{figure}
\centering
\includegraphics[width=9cm]{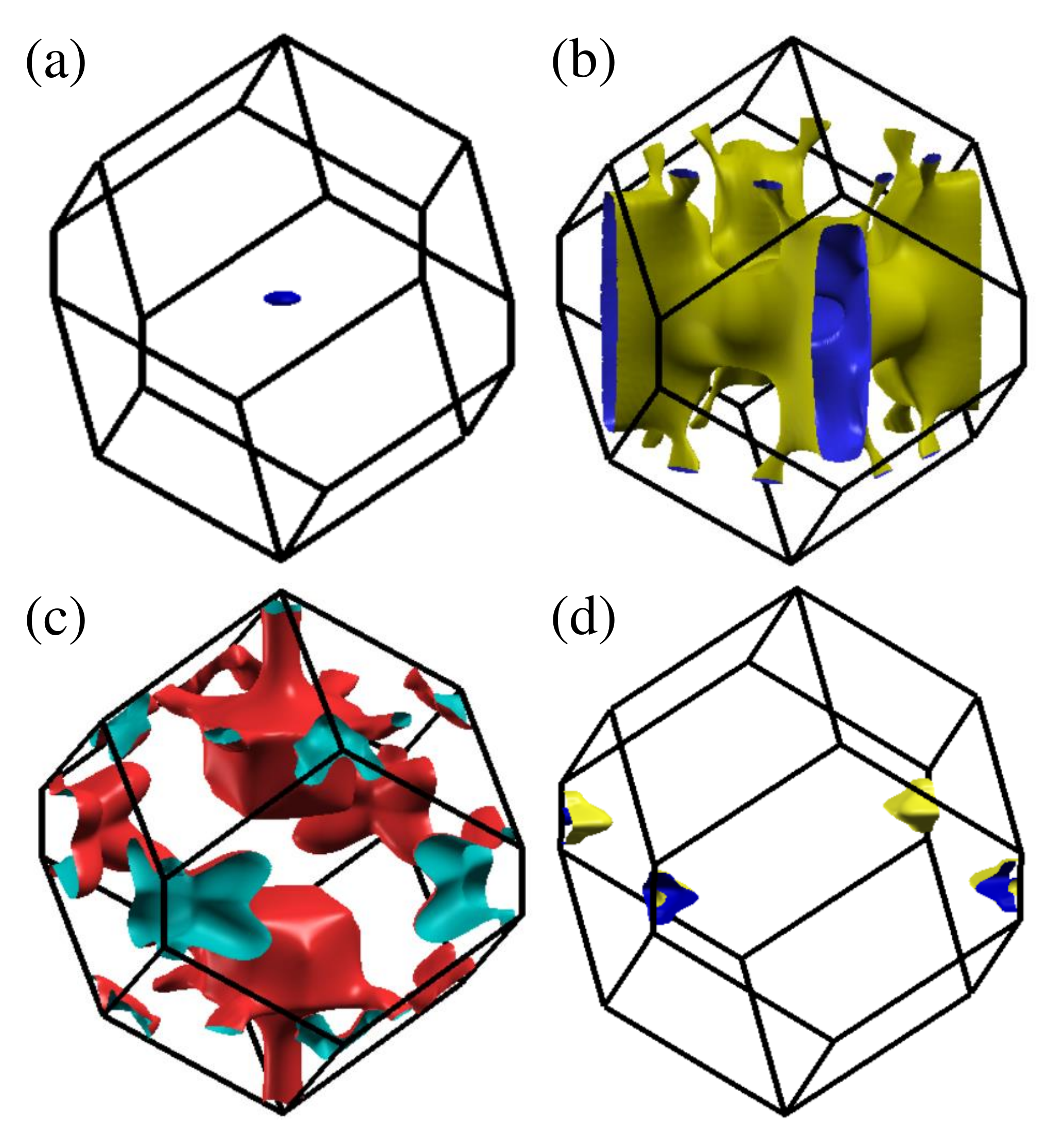}
\caption{\label{fig:fig-4} The calculated Fermi surface sheet with SOC for  Zr$_{2}$Ir}
\end{figure}

To analyze bonding characters between the Ir and Zr atoms, Fig. \ref{fig:fig-3}  is obliquely plotted on the  plane cutting through the  Zr and Ir atoms. In this plane, a region of high charge density  localized around the Zr and Ir atoms gives rise to the metallic character. Furthermore, we have investigated Fermi Surface sheets with the inclusion of SOC as indicated in Fig. \ref{fig:fig-4}. As can be seen from Fig. \ref{fig:fig-4}(a), there is a simple nesting around the $\Gamma$ high symmetry point.  In Figs. \ref{fig:fig-4}(b) and (c) there are complex (electron and hole) sheets for every high high symmetry point except $\Gamma$. In addition, a closed surface is around the $Z_1$ and $M$ as seen in Figs. \ref{fig:fig-4}(c) and (d).

\subsection{Mechanical properties}

Elastic constants are used to describe information about the mechanical and dynamic properties of solids~\cite{Tay-19, Tay-21}. There are six independent single-crystal elastic constants for body-centered tetragonal crystal structure, given as $C_{11}$, $C_{12}$, $C_{13}$, $C_{33}$, $C_{44}$ and $C_{66}$. Although our calculated elastic constant values with and without SOC are presented in Table \ref{tab:table-2} along with other available theoretical results~\cite{Wu-17} that are in agreement, we could not compare them because there are no experimental data. Extensively, for the body-centered tetragonal structure, single-crystal elastic constants of Zr$_{2}$Ir should satisfy the following the mechanical stability criteria~\cite{Wu-07}:

\begin{eqnarray}
C_{11}>0, C_{33}>0, C_{44}>0, C_{66}>0 \nonumber\\ (C_{11}-C_{12})>0,  (C_{11}+C_{33}-2C_{13})>0 \nonumber\\
\bigg{[}2(C_{11}+C_{12})+C_{33}+4C_{13}\bigg{]}>0
\end{eqnarray}

\noindent  The calculated elastic constants accordingly meet the mechanical stability criteria, which indicates that the Zr$_{2}$Ir compound is mechanically stable. The elastic constants $C_{11}$ and $C_{33}$ determine the resistance to linear compression under tension along the x-axis and z-axis, respectively.  The value of $C_{11}$ is higher than that of $C_{33}$ when the inclusion of SOC. It means that the z-axis is more compressible than the x-axis. However, as shown in Table \ref{tab:table-1}, the value of $C_{33}$ is higher than $C_{11}$ without SOC, which indicates the opposite situation. Another remarkable result that $C_{11}$ and $C_{33}$ values are considerably higher than $C_{44}$ and $C_{66}$ values. Therefore, unidirectional compression has a stronger resistance than shear deformation learned from $C_{44}$ and $C_{66}$. Furthermore, the bulk modulus ($B$) and shear modulus ($G$) can be estimated from single-crystal elastic stiffness constants by using Voigt, Reuss, and Hill (VRH) approximations~\cite{Hill-52}. While the Voigt approach~\cite{Voi-28} represents the maximum limit of the elastic modulus, the Reuss approach~\cite{Reu-29} defines the minimum limit. The average value between the Voigt and Reuss boundaries is expressed by Voigt-Reuss-Hill approximations~\cite{Hill-52}. These elastic moduli can be given by the following formula~\cite{Wu-07}:

\begin{eqnarray}
B_{V}=\frac{1} {9}(2(C_{11}+C_{12})+C_{33}+4C_{13})\nonumber\\
G_{V}=\frac{1} {9} (M+3C_{11}-3C_{12}+12C_{44}+6C_{66})\nonumber\\
B_{R}=C^2/M\nonumber\\
G_{R}=15(18B_{V}/C^2)+6/(C_{11}-C_{12})+\nonumber\\
6/C_{44}+(3/C_{66})^{-1}\nonumber\\
B_{H}=(B_{V}+B_{R})/2\nonumber\\
G_{H}=(G_{V}+G_{R})/2\nonumber\\
M=C_{11}+C_{12}+2C_{33}+4C_{13}\nonumber\\
C^2=(C_{11}+C_{12})C_{33}-2C_{11}^2
\end{eqnarray}

\noindent  where $B_{V}$, $B_{R}$, $G_{V}$ and $G_{R}$ are determined as Voight bulk,  Reuss bulk,  Voight shear,   and   Reuss shear, respectively. $B_{H}$ and $G_{H}$ are also Voigt-Reuss-Hill bulk  modulus and shear modulus. Our calculated results with and without the inclusion of SOC are listed in Table \ref{tab:table-1}. Although there is no experimental data, there are theoretical values~\cite{Wu-07, Jai-13} to be compared with our results given in Table \ref{tab:table-2}. Our calculated values are consistent with the available data. The elastic modulus such as bulk modulus, Young's modulus, and shear modulus also provide information about the hardness and strength of materials. It is known that the greater the elastic modulus, the harder the material.  Besides these, Young's modulus $E$ and Poisson's ratio ($v$) can be found using $B_{H}$ and $G_{H}$~\cite{Wu-07}:

\begin{eqnarray}
~E=\frac{9B_{H}G_{H}}{3B_{H}+G_{H}}~,~~~~~~~~v=\frac{3B_{H}-2G_{H}}{2(3B_{H}+G_{H})}
\end{eqnarray}

Young's modulus $E$ is used to measure stiffness material. Our calculated values for Young's modulus are calculated as 85.52 GPa and 86.68 GPa with and without SOC, which are close to the previous result~\cite{Wu-17} as given in Table \ref{tab:table-3}.  The Poisson ratio gives information about the bonding character of materials and the brittle/ductile nature of a material. From the point of view of the bonding character, if the Poisson ratio is around 0.33, it indicates that there is a metallic bond in the material~\cite{Lev-09, Dar-19, Hai-01}. Our results are found to be 0.39 with and without SOC showing a metallic bond in Zr$_{2}$Ir compound. Accordingly, our calculated Poisson’s ratio is close to the the previous result (0.405)~\cite{Wu-17} as seen Table \ref{tab:table-3}. Examining the Poisson ratio ($v$) in terms of the brittle/ductile nature of a material, if $v$ is greater than 0.26, the material will behave brittle, otherwise it will be ductile~\cite{Hai-01, Fra-83}. According to this, the compound Zr$_{2}$Ir represents ductile behavior because the value of $v$ is found 0.39 which is greater than 0.26. Apart from Poisson's ratio $v$, Pugh criteria ($B_H$/$G_H$) and Cauchy pressure ($C_{12}$-$C_{66}$ and $C_{13}$-$C_{44}$)  are also used to obtain information about the brittle/ductile structure of a material. According to Pugh criteria, if the $B_H$/$G_H$ ratio is larger than the critical value of  1.75, the materials are ductile, otherwise, the materials show brittle properties~\cite{Pug-54}. Our calculated values are found to be 4.35 and 4.34 with and without SOC as listed in Table \ref{tab:table-3}.  According to our results with and without SOC as seen in Table \ref{tab:table-3},  Zr$_{2}$Ir is ductile. There are also two Cauchy pressure conditions for the tetragonal crystal
system which are  $C_{12}$-$C_{66}$ and $C_{13}$-$C_{44}$~\cite{Far-94}. When the value
of these conditions is positive, the material is ductile; if
negative, the material is brittle. Zr$_{2}$Ir is ductile because
these results are positive. It is found that Zr$_{2}$Ir shows ductility in  Poisson ratio, Pugh criteria, and Cauchy pressure.

The Debye temperature ($\Theta_D$) is an important physical value and is closely related to thermodynamic properties such as thermal conductivity, isothermal compressibility, melting point, coefficient of thermal expansion, and specific heat~\cite{Ina-83, Pang-11, Toh-06}. It also gives information about the high hardness and melting temperature of materials. Here, the  value of  $\Theta_D$  for Zr$_{2}$Ir  is
estimated from the average  sound  velocity ($V_M$) by the following formula~\cite{And-63}:

\begin{table*}
\centering
\caption{\label{tab:table-2}  Calculated elastic   constants  ($C_{11}$,  $C_{12}$, and $C_{44}$ in  GPa).}

\small{\begin{tabular}{c|cccccc} \hline
                &$C_{11}$ & $C_{12}$ & $C_{13}$ &  $C_{33}$ &  $C_{44}$ &  $C_{66}$ \\\hline\hline
Zr$_{2}$Ir (w/) & 186.11  & 163.85  & 83.95 & 183.84 & 34.64  & 33.63 \\
Zr$_{2}$Ir (w/o)& 186.81 & 163.44    &83.30     &195.06  & 33.25 &33.76\\
Zr$_{2}$Ir (w/o) Theo.~\cite{Wu-17}& 181.80  & 171.80    &84.10     &190.50  & 41.30 &47.80
\\\hline
\end{tabular}}
\end{table*}

\begin{table*}
\centering
\caption{\label{tab:table-3}   Voight bulk modulus ($B_V$), Reuss bulk modulus ($B_R$), average of the Voight  and Reuss bulk modulus ($B_H$) (in GPa);  Voight shear modulus ($G_V$), Reuss shear modulus ($G_R$), an average of the Voight and Reuss  shear modulus ($G_H$) (in GPa); Young's modulus E  (in GPa),  $B_H$/$G_H$ ratio,  Poisson's ratio  $v$,
transverse velocity $V_T$ (in m/s), longitudinal velocity $V_L$ (in m/s), average velocity $V_M$ (in m/s) and temperature $\Theta_D$  (in K) for Ti$_{3}$Sb.}
\small{\begin{tabular}{c|ccccccccccccc} \hline
                 &$B_V$~& $B_R$  & $B_H$  &  $G_V$ &  $G_R$ &  $G_H$ &  $E$   & $B_H/G_H$ &  $v$ & $V_T$  &$V_L$   & $V_M$  & $\Theta_D$\\\hline\hline
Zr$_{2}$Ir (w/)     &135.51  &131.58    &133.54   &33.54    &25.84&30.69     & 85.52   & 4.35 & 0.39 &1742&4152&1970&214\\
Zr$_{2}$Ir (w/o)  &136.53  & 133.71   &  135.12 & 35.96  & 26.26 & 31.11 & 86.68  & 4.34  &0.39  &1757&4178&1987&216  \\\hline
  Theo.~\cite{Wu-17}  &  &    &138.50  &   &  & 28 & 78.60  & 4.95  &0.405 &   &   &  & $\simeq$ 226\\
Exp.~\cite{Fis-74}   & & & & & &     &    &  & & & &&  200 \\
Exp.~\cite{Man-21}   & & & & & &     &    &  & & & && 209 \\\hline
\end{tabular}}
\end{table*}

\begin{eqnarray}
\Theta_D=\frac{h}{k_B}\bigg{(}\frac{3n}{4\pi}\frac{N_A\rho}{M}\bigg{)}^{1/3}V_M
\end{eqnarray}

\noindent where $h$ is the Planck and $k_B$ is the Boltzmann's
constants, $n$ is the concentration, $N_A$ is the
Avogadro number,  $\rho$ is the  mass density and  M is  the
molecular  weight  for the compound. Therefore, $V_M$ is evaluated
by the following equation~\cite{And-63};\\

\begin{eqnarray}
V_M=\bigg{[}\bigg{(}\frac{1}{3}\bigg{(}\frac          {2}{V_T^3}+\frac
  {1}{V_L^3}\bigg{)}\bigg{]}^{-1/3},
\end{eqnarray}

\noindent where $V_T$ and $V_L$ represent transverse and longitudinal sound
velocities. They are found using $B_H$ and $G_H$  as follows:

\begin{eqnarray}
V_L=\bigg{(}\frac{3B_H+4G_H}{3\rho}\bigg{)}^{1/2},
V_T=\bigg{(}\frac{G_H}{\rho}\bigg{)}^{1/2}
\end{eqnarray}

The calculated value of Debye temperature ($\Theta_D$) is 214 K with SOC and 216 K without SOC as listed in Table \ref{tab:table-3}. It is seen that the value of Debye temperature ($\Theta_D$) decreases when SOC is included.  In addition, Debye temperature ($\Theta_D$) of Zr$_{2}$Ir in existing experimental and theoretical studies are given in Table \ref{tab:table-3}. While our calculated Debye temperature ($\Theta_D$) is higher than the experimental values~\cite{Man-21, Fis-74}, it is lower than the available theoretical value~\cite{Wu-17}. These values will become more meaningful if we compare them with other intermetallic compounds so that they do not remain lean. For example,
the Debye temperatures for Zr$_{2}$Rh, Zr$_{2}$Ir, Zr$_{2}$Co, and Zr$_{2}$Ni compounds are given as 188 K, 200 K, 180 K, and 221 K, respectively~\cite{Fis-74}. They are also crystallized in the tetragonal C16 structure like Zr$_{2}$Ir. When our calculated Debye temperatures for Zr$_{2}$Ir with and without soc are compared with those of other compounds, that of Zr$_{2}$Ni is again the highest.

\subsection{Phonon dispersion and electron--phonon coupling}

\begin{figure*}
\bc
\includegraphics[width=16.75cm]{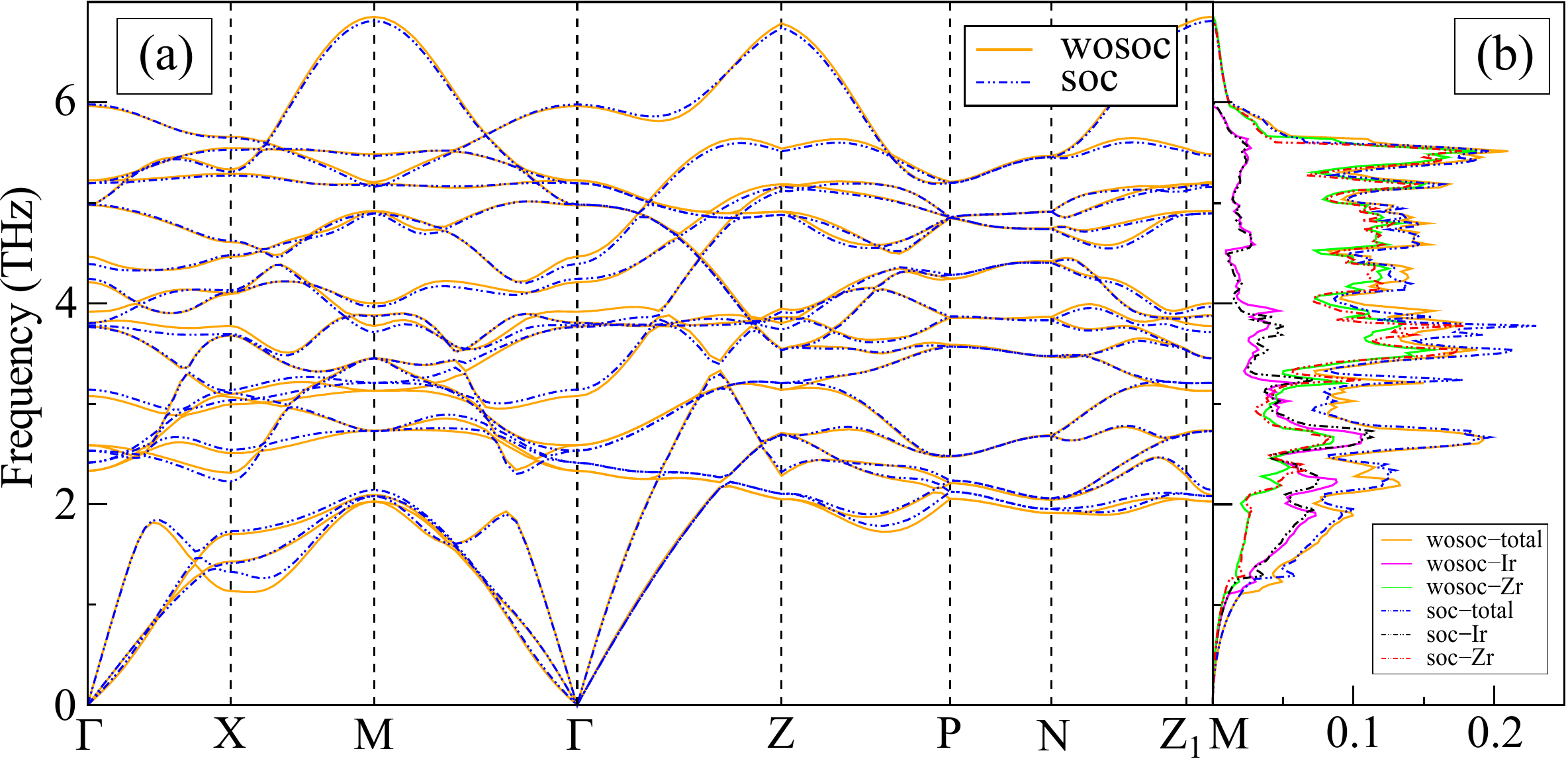}
\ec
\caption{\label{fig:fig-5} The calculated (a) phonon dispersion curves and (b) total and partial phonon density of states of Zr$_{2}$Ir with and without the inclusion of SOC.}
\end{figure*}

Next, we investigate the phonon distributions of Zr$_{2}$Ir. The primitive cell of the compound Zr$_{2}$Ir consists of six atoms, so there are a total of eighteen phonon branches that arise.  Among these three are acoustic and the rest are fifteen optical branches. The calculated acoustic and optical branches along the high symmetry points in the Brillouin region with and without SOC are plotted in Fig. \ref{fig:fig-5}(a). As seen in Fig. \ref{fig:fig-5}(a), Zr$_{2}$Ir compound is dynamically stable because there are no negative frequencies. There is a gap between acoustic and optical branches along with the $\Gamma$-$X$-$M$-$\Gamma$ high symmetry points. The crossover of acoustic and optical modes are also seen in $\Gamma$-$Z$-$P$-$N$-$Z_1$-$M$ high symmetry points
in the Brillouin zone. This is due to the mass ratio between Zr and Ir. In addition, the highest frequencies of the optical mode are very close at 7 THz. The lowest optical phonon branches are found between $Z$ and $P$ points while the highest optical phonon branches are found at $M$ point. Furthermore, it is seen that the highest acoustic modes at between $\Gamma$ and $Z$ points.  We have also presented total and partial phonon density of states (PhDOS) as given in Fig. \ref{fig:fig-5}(b) with and without SOC. We have found that the Zr atom mainly contributes to the high frequencies while the Ir atom dominates the low frequencies region. This is because Ir has a relatively heavier atomic mass.

\begin{figure}
\centering
\includegraphics[width=8.5cm]{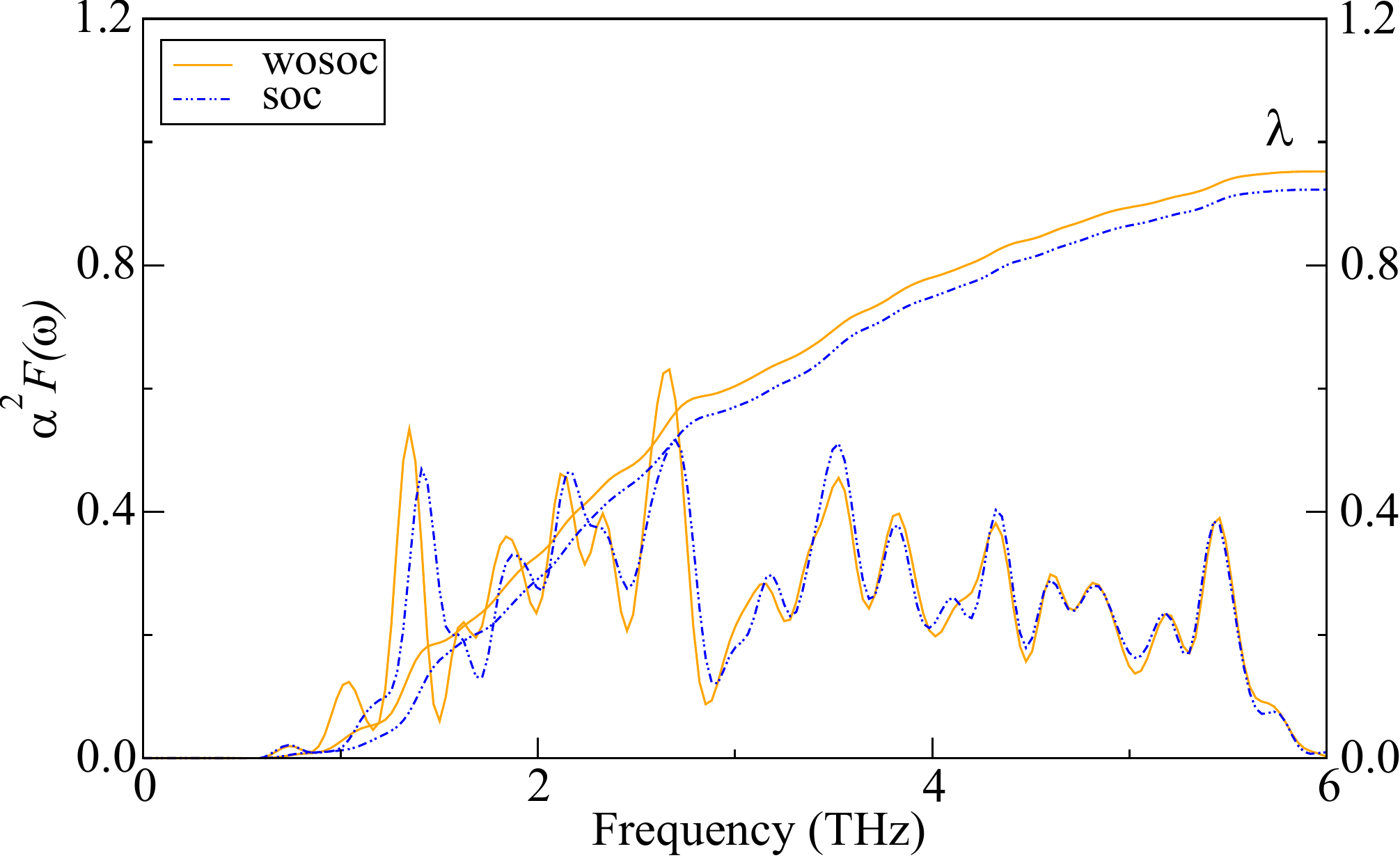}
\caption{\label{fig:fig-6} Eliashberg spectral function $\alpha^2F(\omega)$ and integrated electron-phonon coupling constant $\lambda$ of Zr$_{2}$Ir.}
\end{figure}

To predict the superconducting critical temperature  ($T_c$), Mc--Millan’s equation modified  by  Allen and  Dynes is employed~\cite{Mil-68,  All-75a, All-75b}. According to this transition temperature ($T_c$) can be given following equation:

\begin{eqnarray}
T_c=\frac{\omega_{log}}{1.2}exp\Big(
-\frac{1.04(1+\lambda)}{\lambda-\mu^{\star}(1+0.62\lambda)}
\Big)
\end{eqnarray}

\noindent
where the dimensionless parameter $\mu^{\star}$ represents the effective Coulomb repulsion which  is generally  taken  from 0.10  to  0.13. The parameters $\lambda$ and $\omega_{log}$ are also logarithmically averaged frequency and the  electron-phonon coupling constant. They are evaluated by

\begin{eqnarray}
\lambda=2\int\frac{\alpha^2F(\omega)}{\omega}d\omega \\
ln\omega_{log}=\frac{2}{\lambda}\int
d\omega\frac{\alpha^2F(\omega)}{\omega}ln\omega\
\end{eqnarray}

\noindent where the parameter $\alpha^2F(\omega)$ is Eliashberg spectral function. In Fig. \ref{fig:fig-6}, $\alpha^2F(\omega)$ and integrated electron-phonon coupling constant $\lambda$ with  respect to frequency are calculated when the effective Coulomb repulsion is taken as  $\mu^{\star}$=0.10. The calculated logarithmically averaged frequency $\omega_{log}$ is 122 K with SOC and 118 K without SOC.  The electron-phonon coupling constant is  0.93 with SOC and 0.96 without SOC, listed in Table \ref{tab:table-4}, which is higher than the experimental value (0.83)~\cite{Man-21}. The superconducting critical temperature $T_c$  is calculated to be  7.50 K with SOC and 7.62 K without SOC.  It  is in good agreement with the experimental results given as 7.3 K, 7.4 K, and 7.5 K~\cite{Mc-71, Man-21, Fis-74}, respectively.  $T_c$   gets closer to the experimental values with the inclusion of SOC. This trend is also supported by the electron-phonon coupling constant by the inclusion of SOC. We have also seen a very small decrease in $N({E_F})$ with the inclusion of SOC. These results are indicated that SOC plays a role in the superconductivity properties of Zr$_{2}$Ir material. We have further analyzed the specific heat coefficient ($\gamma_N$) by the following equation;

\begin{eqnarray}
N(E_F)=\frac{3\gamma_N}{\pi^2k_B^2(1+\lambda_{ep})}
\end{eqnarray}

\noindent where we can derive $\gamma_N$ as $\gamma_N$=(1/3)($\pi^2$$k_B^2$$N(E_F))$(1+$\lambda$).  It is found to be 14.21 ~mJ/mol~K$^2$  with  and without  SOC for Zr$_{2}$Ir compounds. It is also smaller than the experimental values of 17.42~mJ/mol~K$^2$ and 17.00~mJ/mol~K$^2$~\cite{Man-21, Fis-74}.

\begin{table}
\centering
\caption{\label{tab:table-4}
  $\lambda$,  $\omega_{log}$ (in K), $N(E_{F})$ (in States/eV), $T_c$   (in  K) and $\gamma_N$ (in mJ/mol~K$^2$)
  with  or  without  SOC  for   the  Zr$_{2}$Ir
  compounds. For $T_c$ calculation, $\mu^{\star}$= 0.10 is taken.}
\begin{tabular}{c|c|c|c|c|c|c|}
\hline
                 &~ $\lambda$ ~& ~$\omega_{log}$~ & ~$N(E_F)$~&~ $T_c$~& $\gamma_N$\\\hline\hline
Zr$_{2}$Ir (w/)  & 0.93      & 122  &  6.25  & 7.50  & 14.21 \\\hline
Zr$_{2}$Ir (w/o) & 0.96      & 118  &  6.26  & 7.62  & 14.21  \\\hline\hline
Zr$_{2}$Ir Exp.~\cite{Mc-71} &       &        &       &  7.30  &  \\\hline
Zr$_{2}$Ir Exp.~\cite{Fis-74} &       &        &       &  7.50  & 17.00  \\\hline
Zr$_{2}$Ir Exp.~\cite{Man-21} & 0.83      &        &       &  7.40  & 17.42
\\\hline
\end{tabular}
\end{table}

Our calculated parameters $N(E_F)$, $\omega_{log}$ and $\lambda$ are interrelated, which are very essential in critical temperature estimation.  For example, the greatest contribution of $\lambda$  comes mainly from Ir atoms in the low-frequency regime up to about 3 THz, which has about 60\% of the value for $\lambda$  in this region. It is seen in our calculation and figures that the value of 0.93 (0.96) with (without) SOC reaches the rest of $\lambda$, which is caused by the Zr atom. This result also explains the strong electron-phonon interaction, meaning that $\lambda$ is greater than the weak electron coupling regime value of 0.5. When we go back to the electronic structure calculation, although a significant contribution of the d orbital of  Zr atom was observed at and around the Fermi level, no difference could be detected in the SOC effect compared to the case without SOC. It also seems that SOC does not have much effect on $\lambda$  either. These generally did not have large splits in the bands under the SOC effect. This is probably due to the relatively weak SOC strength of the Ir or Zr atoms.

\section{Summary}

In this work, we have comprehensively presented the structural, electronic, mechanical, phononic, and superconducting properties using density functional theory with and without SOC. Our density functional theory derived results confirmed that Zr$_{2}$Ir compound is mechanically and dynamically stable.  The electronic band structure demonstrates that Zr$_{2}$Ir has a metallic character. The bulk modulus, shear modulus, and Young modulus have also been derived from our calculated elastic constants. The Zr$_{2}$Ir compound is predicted to be ductile according to Poisson’s ratio, Pugh criteria, and Cauchy pressure. The critical temperature  ($T_c$), electron-phonon coupling constant $\lambda$, logarithmically averaged frequency $\omega_{log}$, and electronic normal state coefficient ($\gamma_N$)  were computed with and without the inclusion of SOC when $\mu^{\star}$ was taken 0.10. The calculated superconducting critical temperature $T_c$ of Zr$_{2}$Ir is 7.50 K (7.62 K) with (without) SOC. It has been found that the superconductivity properties slightly decrease with SOC and get closer to the experimental value. This study is expected to provide some information about the Ir-Zr binary system.

\section{Acknowledgments}

This study was computationally supported by by T. C. Strategy and Budget Directorate under
Project No: 2019K12-92587.

\section*{AUTHOR DECLARATIONS}
\textbf{Conflict of interest}
The authors have no conflicts to disclose.

\section*{DATA AVAILABILITY STATEMENTS}
Data is available on request from the authors.

\addcontentsline{toc}{section}{\protect\numberline{}{Bibliography}}


\begin{thebibliography}{98}

\bibitem{Mit-99}  Y. Yamabe-Mitari, Y. Ro, T. Maruko, H. Harada,  Microstructure dependence of the strength of Ir-base refractory superalloys, Intermetallics 7 (1999) 49, https://doi.org/10.1016/S0966-9795(98)00010-7.

\bibitem{Wu-17}  Jun. Wu, B. Zhang, Y. Zhan,  Ab initio investigation into the structure and properties of Ir–Zr intermetallics for high-temperature structural applications,  Comput. Mater. Sci. 131 (2017) 146–159, https://doi.org/10.1016/j.commatsci.2017.01.047.

\bibitem{Cor-03}   L. Cornish, B. Fischer, R. V\"{o}lkl, Development of Platinum-Group-Metal Superalloys for High-Temperature Use,  MRS. Bull. 28 (2003) 632-638, https://doi.org/10.1557/mrs2003.190.

\bibitem{Che-03} K. Chen, L. Zhao, J.S. Tse,   Ab initio study of elastic properties of Ir and Ir$_{3}$X compounds, J. Appl. Phys. 93 (2003) 2414, https://doi.org/10.1063/1.1540742.

\bibitem{Ran-06} H. Ran, Z. Du, Thermodynamic assessment of the Ir–Zr system, J. Alloys. Compd. 413 (2006) 101–105, https://doi.org/10.1016/j.jallcom.2005.06.060.

\bibitem{Mit-05} Y. Yamabe-Mitarai, T. Maruko, T. Miyazawa, T. Morino,  Solid solution
hardening effect of Ir, Mater. Sci. Forum 475-479 (2005) 703–706,
https://doi.org/10.4028/www.scientific.net/MSF.475-479.703.

\bibitem{Sha-06}  J.B. Sha, Y. Yamabe-Mitarai, Saturated solid-solution hardening behaviour of Ir–Hf–Nb refractory superalloys for ultra-high temperature applications,
Scripta. Mater. 54 (2006) 115–119, https://doi.org/10.1016/j.scriptamat.2005.08.038.

\bibitem{Zhan-18} M. Zhang,  R. Cao, M. Zhao,  J. Du, K. Cheng,    Unexpected Ground-State Structure and Mechanical Properties of Ir$_2$Zr Intermetallic Compound, Mater. 11 (1) (2018) 103, https://doi.org/10.3390/ma11010103.

\bibitem{Kup-74} V.V. Kuprina, G.I. Kuryachava,  Vestn. MGU, Ser. 2, Khim. 15:371–373 ;transl: Moscow Univ Chem Bull 29 (1974) 88,

\bibitem{Ere-74}   V.N. Eremenko, T.D. Shtepa, E.L. Semenova, in:  Rykhal R,
L’vov M (Eds.), Tszisy Dokl. Vses. Konf. Kristallokhim. Intermet.
Soedin., 2nd ed., Inst Probl Materialoved, Kiev, 1974, 28.


\bibitem{Ere-78}  V. N. Eremenko, E.L. Semenova, T.D. Shtepa, Influence of Rhodium, Iridium and Osmium on the Polymorphic Transformation in Zirconium, Russ. Metall. 2 (1978)
158-160.

\bibitem{Erem-78} V. N. Eremenko, E.L. Semenova, T.D. Shtepa, Yu.V. Kudryavtsev,  X-Ray Study of the Zr-Rh and Zr-Ir Base Phases at High Temperatures.
Dop. Akad. Nauk. Ukr. RSR, A, Fiz.–Mat. Tekn. 10 (1978) 943-945.

\bibitem{Oka-92} H. Okamoto,   The Ir-Zr (Iridium-Zirconium) System. Journal of Phase Equilibria  13 (1992) 653 - 656.

\bibitem{Ere-80} Eremenko, V.N.; Semenova, E.L.; Shtepa,  State diagram of the Zr-Ir system, Russ. Metall. 5 (1980) 210–213.

\bibitem{Mc-71} S. McCarthy, The superconductivity and magnetic susceptibility of some zirconium-transition-metal compounds; evidence for an anticorrelation, J. Low. Temp. Phys. 4 (1971) 489-501, https://doi.org/10.1007/BF00631128.

\bibitem{Fis-74} Z. Fisk, R. Viswanathan, G.W. Webb,  The Relation between normal state properties and T$_C$ for some Zr$_2$X Compounds, Solid State Commun. 15 (1974) 1797—1799, https://doi.org/10.1016/0038-1098(74)90089-1.

\bibitem{Fag-06} R. L. Fagaly,   Superconducting quantum interference
device instruments and applications, Rev. Sci. Instru.
77 (2006)  101101, https://doi.org/10.1063/1.2354545

\bibitem{Kas-21} Md. Riad Kasem, Aichi Yamashita, Yosuke Goto, Tatsuma D. Matsuda and Yoshikazu Mizuguchi,  Synthesis of high-entropy-alloy-type superconductors (Fe,Co,Ni,Rh,Ir)Zr$_2$ with tunable transition temperature, J. Mater. Sci.  56 (2021) 9499–9505, https://doi.org/10.1007/s10853-021-05921-2.

\bibitem{Man-21}  M. Mandal, C. Patra, A. Kataria, D. Singh, P. K. Biswas, J. S. Lord, A. D. Hillier, and R. P. Singh,  Superconducting ground state of the nonsymmorphic superconducting compound Zr$_2$Ir, Phys. Rev. B 104 (2021) 054509, https://doi.org/10.1103/PhysRevB.104.054509.

\bibitem{Gia-09} P. Giannozzi et al.,  QUANTUM ESPRESSO: a modular and open-source software project for quantum simulations of materials, J. Phys.: Condens. Matter. 21 (2009) 395502, https://doi.org/10.1088/0953-8984/21/39/395502.
	
\bibitem{Gia-17}P. Giannozzi et al.,  Advanced capabilities for materials modelling with Quantum ESPRESSO, J.  Phys. Condens. Matter. 29 (2017) 465901, https://doi.org/10.1088/1361-648X/aa8f79.

\bibitem{Blo-94} P. E. Bl\"{o}chl, Projector augmented-wave method,
Phys. Rev. B 50 (1994) 17953, https://doi.org/10.1103/PhysRevB.50.17953

\bibitem{Per-96}   J. P. Perdew, K. Burke, M. Ernzerhof, Generalized Gradient Approximation Made Simple, Phys Rev Lett 77 (1996) 3865-3868, https://doi.org/10.1103/PhysRevLett.77.3865; Erratum,  Phys Rev Lett 78 (1997) 1396,https://doi.org/10.1103/PhysRevLett.78.1396.

\bibitem{Fis-92} T. H. Fischer, J. Almlof, General Methods for Geometry and Wave Function Optimization, J. Phys. Chem. 96 (1992) 9768-9774, https://doi.org/10.1021/j100203a036.

\bibitem{Met-89}    M. Methfessel, A. T. Paxton, High-precision sampling for Brillouin-zone integration in metals,  Phys. Rev. B 40 (1989) 3616,  https://doi.org/10.1103/PhysRevB.40.3616.

\bibitem{Mon-76} H. J. Monkhorst, J. D. Pack,  Special Points for Brillouin-Zone Integrations, Phys. Rev. B 13 (1976) 5188--5192, https://doi.org/10.1103/PhysRevB.13.5188.

\bibitem{Cor-16} A. D. Corso, Elastic constants of beryllium: a first-principles investigation, J. Phys.  Condens.  Matter. 28 (2016) 075401, https://doi.org/10.1088/0953-8984/28/7/075401.

\bibitem{qe} https://dalcorso.github.io/thermo$_{-}$pw/

\bibitem{Mig-58}  A. B. Migdal, Interaction Between Electrons and Lattice Vibrations in a Normal Metal,  Sov. Phys. JETP. 34 (1958) 996.

\bibitem{Eli-60}    G. M.  Eliashberg,  Interaction Between Electrons and Lattice Vibrations in a Superconductor. Zh  Eksp  Teor  Fiz 38 (1960) 966,  English trans, Sov  Phys JETP 11 (1960) 696.

\bibitem{Mil-68}  W. L. McMillan,  Transition Temperature of Strong-Coupled Superconductors, Phys. Rev. 167 (1968) 331--344, https://doi.org/10.1103/PhysRev.167.331.

\bibitem{All-75a}   P. B . Allen, R.C. Dynes,   Transition temperature of strong-coupled superconductors reanalyzed, Phys. Rev. B 12 (1975)  905-922, https://doi.org/10.1103/PhysRevB.12.905

\bibitem{All-75b} P. B . Allen,  R.C. Dynes, Superconductivity at very strong coupling, J. Phys. C. 8 (9) (1975) L158-L163, https://doi.org/10.1088/0022-3719/8/9/020.

\bibitem{Mom-11}   K.  Momma, F.   Izumi,  VESTA 3 for three-dimensional visualization of crystal, volumetric and morphology data,  J. Appl. Crystallogr. 44 (2011) 1272,  https://doi.org/10.1107/S0021889811038970.


\bibitem{Kok-99} A. Kokalj,  XCrySDen—a new program for displaying crystalline structures and electron densities, J. Mol. Graph. Model. 17 (3-4) (1999) 176–179, https://doi.org/10.1016/S1093-3263(99)00028-5

\bibitem{Jai-13}   A. Jain, S. P. Ong , G. Hautier, W. Chen,  W. D. Richards, S. Dacek, S. Cholia, D. Gunter, D. Skinner, G. Ceder, K. A. Persson, Commentary: The Materials Project: A materials genome approach to accelerating materials innovation, APL  Materials  1  011002, (2013)  https://doi.org/10.1063/1.4812323

\bibitem{Tay-19} C. Tayran, M. ~\c{C}akmak, Electronic structure, phonon and superconductivity for WP 5d-transition metal, J Appl Phys. 126  (2019)  175103, https://doi.org/10.1063/1.5122795.

\bibitem{Tay-21} C. Tayran, M. ~\c{C}akmak, Electronic, phononic and superconducting properties of trigonal Li$_2$MSi$_2$ (M = Ir, Rh), J. Phys.  Condens. Matter. 33 (2021) 065502, https://doi.org/10.1088/1361-648X/abc405.

\bibitem{Wu-07} Z-J.  Wu, E-J. Zhao, H-P. Xiang, X-F. Hao, X-J. Liu , J. Meng,  Crystal structures and elastic properties of superhard IrN$_2$ and IrN$_3$ from first principles, Phys. Rev. B.  76 (2007) 054115, 10.1103/PhysRevB.76.054115.


\bibitem{Hill-52} R. Hill,  The Elastic Behaviour of a Crystalline Aggregate,  Proc. Phys. Soc.  London 65 (1952) 349, https://doi.org/10.1088/0370-1298/65/5/307.

\bibitem{Voi-28} W. Voigt,  Lehrbuch der Kristallphysik: mit Ausschluss der Kristallopti) (Leipzig: Teubner) (1928).


\bibitem{Reu-29}   A. Reuss,  Berechnung der Fliessgrenze von Mischkristallen auf Grund der Plastizitätsbedingung für Einkristalle. ZAMM—Journal of Applied Mathematics and Mechanics/Zeitschrift für Angewandte Mathematik und Mechanik, 9 (1) (1929) 49-58,  https://doi.org/10.1002/zamm.19290090104

\bibitem{Lev-09} J. B. Levine, S. H. Tolbert, R. B. Kaner,  Advancements in the Search for Superhard Ultra-Incompressible Metal Borides, Adv. Funct. Mater. 19 (2009) 3519, https://doi.org/10.1002/adfm.200901257.

\bibitem{Dar-19}  S. A. Dar, R. Sharma,  A. K. Mishra,  Phonon stability, electronic structure results, mechanical and thermodynamic properties of RbSbO$_3$ and CsSbO$_3$ perovskite oxides: Ab initio investigation, J. Mol. Graph. Model. 90 (2019) 120-127, https://doi.org/10.1016/j.jmgm.2019.04.013.

\bibitem{Hai-01}  J. Haines,  J. M. Leger,  G. Bocquillon, Synthesis and Design of Superhard Materials, Annu. Rev. Mater. Res. 31 (2001) 1-23, https://doi.org/10.1146/annurev.matsci.31.1.1

\bibitem{Fra-83} I. N. Frantsevich,  F. F. Voronov, S. A. Bokuta,
Elastic Constants and Elastic Moduli of Metals and Insulators Handbook, Naukova Dumka,  Kiev (1983), 60–180.

\bibitem{Pug-54}  S. F. Pugh,  XCII. Relations between the elastic moduli and the plastic properties of polycrystalline pure metals, Phil. Mag. 45 (367) (1954) 823-843, https://doi.org/10.1080/14786440808520496.

\bibitem{Far-94} D. Farkas, Interatomic potentials for Ti-Al with and without angular forces, Model Simulat. Mater. Sci. Eng. 2 (1994) 975, https://doi.org/10.1088/0965-0393/2/5/003

\bibitem{Ina-83} H. Inaba, T. Yamamoto, Debye temperature of materials, Netsu Sokutei (1983)  10 (4) 132-14,  https://doi.org/10.11311/jscta1974.10.132

\bibitem{Pang-11}  M. Pang,  Y. Zhan,  W. Jiang,  Y. Du, Ab initio study of AlCu$_2$M (M = Sc, Ti and Cr) ternary compounds under pressures, Comput. Mater. Sci. 50 (10) (2011) 2930-2937, https://doi.org/10.1016/j.commatsci.2011.05.010

\bibitem{Toh-06}  T. Tohei,  A. Kuwabara,  F. Oba,  I. Tanaka,  Debye temperature and stiffness of carbon and boron nitride polymorphs from first principles calculations, Phys. Rev. B.  73 (2006) 064304, https://doi.org/10.1103/PhysRevB.73.064304

\bibitem{And-63}  O. L. Anderson,   A simplified method for calculating the debye temperature from elastic constants, J. Phys.  Chem.  Solids 24 (17) (1963) 909-917, https://doi.org/10.1016/0022-3697(63)90067-2

\end{thebibliography}
\end{document}